\documentclass[conference]{IEEEtran}
\usepackage{amssymb}
\usepackage{amsmath}
\usepackage{amsfonts}
\usepackage{graphicx}
\usepackage{algorithm}
\usepackage{algorithmic}
\usepackage{cite}
\usepackage{epstopdf}

\usepackage{amssymb}
\usepackage{amsmath}
\usepackage{amsfonts}
\usepackage{graphicx}
\usepackage{algorithm}
\usepackage{algorithmic}
\usepackage{cite}
\usepackage{epstopdf}
\usepackage{color}
\usepackage{multirow}
\usepackage{subfigure}
\usepackage{bm}

\begin{document}
\title{ Simplified Multiuser Detection for SCMA with Sum-Product Algorithm}
\author{Kexin Xiao, Baicen Xiao, Shutian Zhang, Zhiyong Chen, and Bin Xia\\
Department of Electronic Engineering, Shanghai Jiao Tong University, Shanghai, P. R. China\\
Email: {\{kexin.xiao, xinzhiniepan, zhangshutian, zhiyongchen, bxia \}@sjtu.edu.cn}
}
\maketitle

\begin{abstract}
Sparse code multiple access (SCMA) is a novel non-orthogonal multiple access technique, which fully exploits the shaping gain of multi-dimensional codewords. However, the lack of simplified multiuser detection algorithm prevents further implementation due to the inherently high computation complexity. In this paper, general SCMA detector algorithms based on Sum-product algorithm are elaborated. Then two improved algorithms are proposed, which simplify the detection structure and curtail exponent operations quantitatively in logarithm domain. Furthermore, to analyze these detection algorithms fairly, we derive theoretical expression of the average mutual information (AMI) of SCMA (SCMA-AMI), and employ a statistical method to calculate SCMA-AMI based specific detection algorithm. Simulation results show that the performance is almost as well as the based message passing algorithm in terms of both BER and AMI while the complexity is significantly decreased, compared to the traditional Max-Log approximation method.

\end{abstract}
\section{Introduction}
For the ability of supporting  massive connections simultaneously, Sparse code multiple access~(SCMA)\cite{nikopour2013sparse}, a  non-Orthogonal Multiple Access (NOMA) scheme, has been regarded as a competitive candidate for Fifth Generation(5G) communication. Commonly, SCMA can be seen as a generalization of sparsely spread CDMA\cite{guo2008multiuser}, with a few numbers of nonzero elements within a large signature length. When multiple layers are multiplexed with unique codebooks\cite{taherzadeh2014scma}, the superposition scheme enable SCMA to benefit from shaping gain. However, to fully acquire these benefits, on the receiver side serious multiple address interference (MAI) is the main difficulty to process multiuser detection. The optimum maximum a posterior (MAP) algorithm obviously shows the best detection performance with considerable calculation complexity.

Meanwhile, some low complexity linear algorithm are proposed to handle this NP-complete problem\cite{verdu1998multiuser} inevitably within some kind of performance loss. Especially, thanks to the sparse structure of SCMA spreading signature, the complex MAP formula\cite{hoshyar2008novel} can be solved step by step within sum-product algorithm or message passing algorithm (MPA)\cite{kschischang2001factor}. Under an overloading system, detection complexity  still increases with the number of users and size of codebooks. Thus, it is essential to take caution to simplify the algorithm and to find a tradeoff between complexity and performance.

In this paper, inspired by the sum-product algorithm, we illustrate and prove how MPA is applied in multiuser SCMA  detection
 and during the process of iteration, we then modify the message passing method to speed up the convergence in a reasonable way, which mainly make prior decisions on these high probability variable in advance. Furthermore, to reduce the calculation of product operation, we covert the whole algorithm into $\mathrm{log}$ domain and present the specific expression of symbols and bits soft message, which can be used in soft channel decoding conveniently. To eliminate the exponent operation, this work proposes two  alternative methods based on the total least square criterion to approximate Jocabian logarithm\cite{viterbi1998intuitive} instead of simple Max-log\cite{zhang2014sparse}. Moreover, inspired by the concept of average mutual information in \cite{xie2012simplified}, we take the SCMA scenario into account for further performance evaluation, proposing the average mutual information of SCMA (SCMA-AMI) and deriving the theoretical expression of SCMA-AMI. Finally, the simulation results show the AMI gap between these simplified methods is negligible. The BER performance under both AWGN and Rayleigh channels are made a comparison between these algorithms, proposed methods present promising results superior to traditional Max-Log method.

For the sake of clarity, the following notational convention unless noted otherwise. The sets of binary and complex numbers are denoted by $\mathbb{B}$ and $\mathbb{C}$. Upper-case calligraphic symbols denote sets, e.g., $\mathcal{X}$. Scalars and vectors are represented by normal, bold fonts respectively. $\mathrm{log}(\cdot)$denotes the natural logarithm operation.

\section{System model}%

\begin{figure*}[t]
            \centering
            \includegraphics[width=7in]{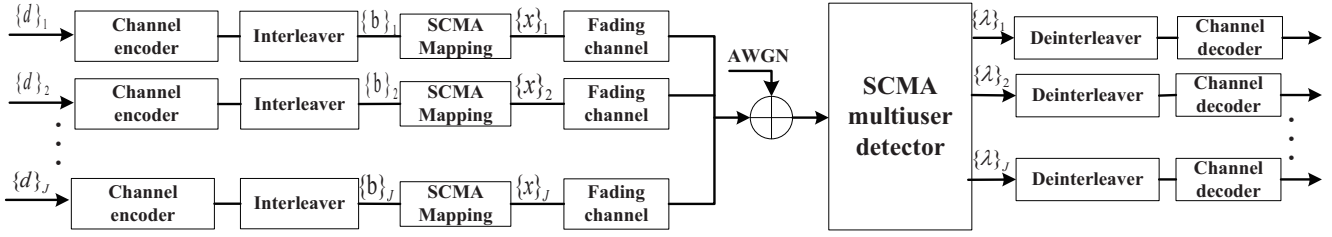}
            \hspace{0.5in}
\caption{Illustration of a SCMA uplink system with J users spreading over K resources elements. }
            \label{SCMA_structure3}
          \end{figure*}

We consider a general uplink SCMA system as shown in Fig.{\ref{SCMA_structure3}} with $J$ users spreading over $K$ orthogonal resources (RE). In order to obtain higher spectrum efficiency, overloading factor defined as $\mu=J/K$ is usually larger than one. On transmitting side, for each user $j$,  $j=1,2...J$, the binary information data stream $\{d\}_j=(d_{1j},...,d_{k_0j})$ are encoded by channel encoder with coding rate $R_j$ then the corresponding output after code bit interleaver is $\{b\}_j=(b_{1j},...,b_{n_0j})$, where $R_j=n_0/k_0$. These encoded bit streams are than mapped into $K$-dimensional complex symbols, under the mapping relationship :
$f:B^{\mathrm{log}_2(M)}\rightarrow \mathcal{X}_j$, $\mathbf{x}_j=f(b_{1j},..., b_{D}), \mathbf{x}_j=(x_{1j},...x_{Kj})$, where  $\mathcal{X}_j$  belongs to $\mathbb{C}^K$ with cardinality $|\mathcal{X}_j|=M$, $ D=\mathrm{log}_2M $. Because of  the sparse feature of SCMA, there are only $N_j<K$ non-zero entries among a $K$ dimensional constellation symbol $\mathbf{x}_j$, the generated codewords are then allocated on orthogonal resources, modulated by  OFDM modulators.
For simplification, here we assume that each layer corresponding to a specific user, transmitting single bit stream. In fact, under prompt scheduling algorithm, more layers can be assigned to different users, while these subtle simpilfication has no essential influence on the subsequent analyzation. During a time slot, on receiver side, the received signal is the superposition of $J$ users' signal plus the ambient noise $\mathbf{n}$, given by
\begin{equation}
 \mathbf{y}=\sum_{i=1}^J\mathrm{diag}({\mathbf{h}_j}){\mathbf{x}_j}+{\mathbf{n}},
\end{equation}
where the received signal vector $\mathbf{y}=(y_1,...,y_K)^T$, $\mathbf{h_j}=(h_{1j},...,h_{Kj})^T$  is the channel vector for user $j$th over $K$ REs, and $\mathbf{n}\sim\mathcal{CN}(0,N_0\mathbf{I})$ is a $K$ dimensional Complex Gaussian noise vector vector.

\begin{figure}[t]
            \centering
            \includegraphics[width=3.2in]{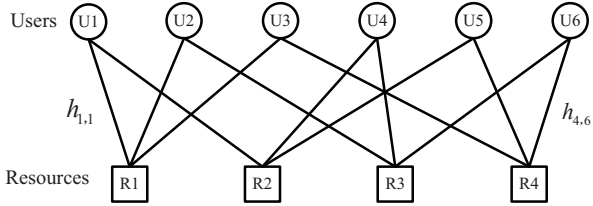}
            \hspace{2in}\parbox{1\linewidth}
{\caption{Factor graph representation of an SCMA system with J=6,K=4,N=2}
\label{factor_graph}}
\end{figure}

The above uplink  multiuser multiplex scheme can be depicted in a bipartite graph $G(\mathcal{V},\mathcal{E})$, which contains $J$ variable nodes $VN$s, $K$ function nodes $FN$s. Besides that, the edge between $VN_j$ and $FN_k$ means that resource k is occupied by user j the nodes. Let $\partial j$ represents the set of FNs connected with the $j$th VN and $\partial k$ has the similar meaning like $\partial j$. In an regular SCMA codebook, the degrees of $VN$s are equal to $\mathbf{d}_v=\partial j=N$,  while the degree of each $FN$ is equal to $\mathbf{d}_f=|\partial k|$ which depends on the overloading factor. In Fig.{\ref{factor_graph}} shows an example of regular factor graph with $J=6$, $K=4$, $N=d_f=2$ and $d_v=3$.  According to the above illustration and formula (1), it is easy to see the interference on $k$th dimension as follows:
\begin{equation}
y_k=\sum_{j\in \partial k}h_{kj}x_{kj}+n_k, k=1,...,K.
\end{equation}
In next section, we will pay attention to the multiuser detection based on the elaborated system model.

\section{SCMA multiuser detection}
In this section, SCMA detector are elaborated in details based on message passing algorithm (MPA). Besides that, we improve and simplify the whole detection process.

\subsection{SCMA Detection with Sum-Product Algorithm}
The optimum detection criterion the object function is given by:
 \begin{equation}
\widehat{ \mathbf{x}}_j=\arg\max_{a\in\mathcal{X}}\sum_{\substack{\{\mathbf{x}_i\}\in \mathcal{X}^J \\\mathbf{x}_j=a}}P(\{\mathbf{x}_i\})\prod_{k=1}^K p(\mathbf{y}_k|\{\mathbf{x}_l\},l\in \partial k),
 \end{equation}
 where$\{\mathbf{x}_i\}$ represent the set of constellation signals of related users. The key point is to effectively deal with the object marginalize product of functions.
 If the factor graph is cycle-free, the factor graph can be converted to an expression tree and each marginal posterior probability of $VN$ can be obtained respectively. Further more, the redundant calculation of each $VN$ can be shared simultaneously in a more effective method, which is proposed as Sum-Product algorithm or Message Passing Algorithm(MPA)\cite{kschischang2001factor} in SCMA.

Consider the bipartite graph depicted in Fig.{\ref{factor_graph}}. During each iteration, messages are first sent from $VN$s to $FN$s; Each $FN$ then computes extrinsic messages and sends back to the $VN$s based on the previously received information. These $VN$-to-$FN$ messages will then be used to calculate the new $FN$-to-$VN$ messages in the next iteration. As $|\mathcal{X}_j|=M$, let $\mathcal{X}_j=\{\mathbf{x}_{j1},...\mathbf{x}_{jm},...,\mathbf{x}_{jM}\}$, where $\mathbf{x}_{jm}=(x_{1jm},...,x_{Kjm})^T$. $\{V_{j\rightarrow k}^{(t)}(\mathbf{x}_{jm})\}$ represent the $m$th codeword message from $VN_j$ to $FN_k$  and $\{U_{k\rightarrow j}^{(t)}(\mathbf{x}_{jm})\}$ represents the message in the reverse direction at the $t $th iteration. These above messages all means the extrinsic message. Let $\partial j/k$ denote the neighborhood of $VN_j$ excluding $FN_k$ and $\partial j/k$ has similar meaning as above. $\mathrm{H}=\{\mathbf{h}_1,...,\mathbf{h}_j,...,\mathbf{h}_J\}$ represents the perfect channel estimation.
The sum and product algorithm for computing the (approximate) posteriori distribution of all layers' symbol are described as follows: in Algorithm 1.

\begin{algorithm}[t]
\caption{Sum-product algorithm for SCMA detection}
\begin{algorithmic}[1]

\STATE \textbf{In put variable}:$\mathbf{y}$, $H$, $\mathcal{X}$
\STATE \textbf{Initial:}
$P(\mathbf{x}_{jm})=1/M$,~$m=1,...,M; j=1,....,J$ \\
      \textbf{for} $j=1,...,J , k=1,...,K$ and m=1,...,M \textbf{do} \\
~~~$V_{j\rightarrow k}^{(0)}(\mathbf{x}_{jm})=0$,~$U_{k\rightarrow j}^{(0)}(\mathbf{x}_{jm})=1$\\

\STATE \textbf{for} $t=1 ~to~ T$  \textbf{do}\\
~~~\textbf{firstly} $VNs\rightarrow FNs$\\
~~~\textbf{for}~all $j$,$k$~and $m$~(if Edge(j,k)~exists)\textbf{do}
\begin{equation}
      V_{j\rightarrow k}^{(t)}(\mathbf{x}_{jm})= P(\mathbf{x}_{jm})\prod\limits_{l\in  \partial j\setminus k}U_{l\rightarrow j}^{(t-1)}(\mathbf{x}_{jm}).
\end{equation}

~~~To keep  numerical stable, here normalize the data:\\

\begin{equation}
     V_{j\rightarrow k}^{(t)}(\mathbf{x}_{jm})=  V_{j\rightarrow k}^{(t)}(\mathbf{x}_{jm})/\sum_{m=1}^M V_{j\rightarrow k}^{(t)}(\mathbf{x}_{jm}).
\end{equation}
~~~\textbf{secondly}~$FNs\rightarrow VNs$\\
~~~\textbf{for} ~all~$k$, $j$and $m$ ~(if Edge(j,k)~exists)~\textbf{do}\\
\begin{align}
U_{k\rightarrow j}^{(t)}(\mathbf{x}_{jm})=\sum_{com}^{|com|} \frac{1}{\pi N_0}exp[-\frac{1}{N_0}(y_k-h_{kj}x_{kjm}  \nonumber\\
-\sum_{i \in\partial k\setminus j}h_{ki}x_{ki\sim}^{(com)})^2] \prod\limits_{i\in\partial k\setminus j}V_{i\rightarrow k}^{(t)}(\mathbf{x}_{jm}),
\end{align}
~~~where $com=...\times \mathcal{X}_i\times ...\times \mathcal{X}_{d_f-1}$ ~is the codewords \\
"$\times$" means Cartesian Product, where$i\in \partial k\setminus j$, $|com|=M^{d_f-1}$\\

\STATE
\textbf{finnal result}:
\textbf{for},$j=1,...,J$; $m=1,...,M$
\begin{equation}
 V_j(\mathbf{x}_{jm})=P(x_{jm})\prod\limits_{k\in \partial j}U_{k\rightarrow j}^{(T)}(x_{jm}).
\end{equation}

\STATE
\textbf{Normalization}
\begin{equation}
     V_{j}(\mathbf{x}_{jm})=  V_{j}(\mathbf{x}_{jm})/\sum_{m=1}^M V_{j}(\mathbf{x}_{jm});~all ~j,m.
\end{equation}

\end{algorithmic}
\end{algorithm}

If the bipartite graph dose not exist any cycles, the iteration number T equals to the max degree of depth otherwise the correct object function can not be obtained, we can only attain an approximative result.

In the final iteration, $\{V_{j}(\mathbf{x}_{jm})\}$ represent the approximate posterior probability of symbol $\mathbf{x}_{jm}$, then each symbols of J layers  can be determined at the same time.
Here we also propose a simplified scheme:
when observing the simulation process, we find that message passing algorithm is an adaptive algorithm, especially under cycled factor graph error detector probability will spread and influence other $VN$ symbol's judgement while some $VN$s' symbol with higher confidence will quickly show the convergent tendency. Thus if in the foremost process of iteration, some $VN$ show an obvious probability distribution, we can reasonable infer the correct transmit constellation on $VN_j^*$.
So we modify and simplify the above sum-product algorithm as follows: in Algorithm 2

\begin{algorithm}[t]
\caption{Modified Sum-product algorithm for SCMA detection}
\begin{algorithmic}[1]

\STATE \textbf{main iteration }:\\
during  the t th iteration\\
for all m and j;\\
\begin{equation}
V_j^{temp}(m)=P(x_{jm})\prod\limits_{k\in \partial j}U_{k\rightarrow j}^{(t)}(x_{jm}),\nonumber
\end{equation} \nonumber
find $m^*=argmax(V_j^{temp}(m))$ in each layer j\\
 $\mathbf{if}$ $max(V_j^{temp}(m^*))>\alpha$  ($\alpha$ can be designed as 0.5)~set
\begin{equation}
V_{j^*}(m)=V_{j^*\rightarrow k}(\mathbf{x}_{j^*m})=\left\{
\begin{aligned}
& 1 &m=m^* \\
&   0 &m\neq m^*\\
\end{aligned}
\right.
\end{equation}
 where$k\in \partial j^*$
\STATE \textbf{other procedure is the same as sum-product algorithm}
when FNs connect the judged $VN_{j^*}$, $com=...\times \mathcal{X}_i\times ...\times \mathcal{X}_{d_f-1}$the cardinality of set com decreases to $|com|=M^{(d_f-2)}$, and there is no need to recalculate  $U_{k\rightarrow j^*}^{(t)}(x_{j^*m})$ any more because of the judgement of $j^*$ in advance.

\end{algorithmic}
\end{algorithm}
In this modified algorithm, if in the t-th iteration the symbol on some ${VN_j^*}$ is judged, when calculating $\{U_{k\rightarrow l}\}$, where $k\in \partial j^*$, ${VN_l}$ share the common ${FN_k}$ with ${VN_j^*}$, the cardinality of set $com$ decreases to $M^{df-2}$, and $\{U_{k\rightarrow j^*}\}$ is no longer to be updated. Thus, the quantity of calculation exponent operations can sharply be curtailed.

\subsection{Simplified SCMA Dectection Based on LLR Type}
To manipulate the Sum-Product Algorithm practicably, the reduction of product calculation is essential. Thus, we utilize log domain to eliminate product operator.

Based on the result in Section III-A: we soon get the following formula: let us fix a reference point $\mathbf{x}_{j1}\in \mathcal{X}_j$, define $\{L_{j\rightarrow k}^{(t)}(\mathbf{x}_{jm})\}$, $\{L_{k\rightarrow j}^{(t)}(\mathbf{x}_{jm})\}$ represent the log domain of $\{V\}$ and $\{U\}$
from formula (4) and (6)
\begin{equation}
\begin{split}
L_{j\rightarrow k}^{(t)}(\mathbf{x}_{jm})&= \mathrm{log}\frac{V_{j\rightarrow k}^{(t)}(\mathbf{x}_{jm})}{V_{j\rightarrow k}^{(t)}(\mathbf{x}_{j1})}\\
&=\mathrm{log}\frac{P(\mathbf{x_{jm}})}{P(\mathbf{x_{j1}})}+\sum\nolimits_{l\in  \partial j\setminus k}L_{l\rightarrow j}^{(t-1)}(\mathbf{x}_{jm}),
\end{split}
\end{equation}

\begin{equation}
\begin{split}
&L_{k\rightarrow j}^{(t)}(\mathbf{x}_{jm})= \mathrm{log}\frac{U_{k\rightarrow j}^{(t)}(\mathbf{x}_{jm})}{U_{k\rightarrow j}^{(t)}(\mathbf{x}_{j1})}=\mathrm{log}\frac{\sum_{n=1}^{|com|}e^{C_{jm}^{(n)}}}{\sum_{n=1}^{|com|} e^{C_{j1}^{(n)}}},
\end{split}
\end{equation}

where \\
\begin{equation}
\begin{split}
C_{jm}^{(n)}=& -\frac{1}{N_0}(y_k-h_{kj}x_{kjm}-\sum\nolimits_{i \in\partial k\setminus j}h_{ki}x_{ki\sim}^{(n)})^2\\+
&\sum\nolimits_{i\in  \partial k\setminus j}L_{i\rightarrow k}^{(t-1)}(\mathbf{x}_{jm}),
\end{split}
\end{equation}
final result
\begin{equation}
\begin{split}
L_{j}(\mathbf{x}_{jm})&= \mathrm{log}\frac{V_{j}(\mathbf{x}_{jm})}{V_{j}(\mathbf{x}_{j1})}\\
&=\mathrm{log}\frac{P(\mathbf{x_{jm}})}{P(\mathbf{x_{j1}})}+\sum\nolimits_{l\in  \partial j\setminus k}L_{l\rightarrow j}^{(T)}(\mathbf{x}_{jm});~all ~m,j.
\end{split}
\end{equation}

Consequently, we proceed further and derive the output bit soft message LLR, as $D=\mathrm{log}_2M$ bits are grouped together and then mapped to a constellation symbol, at the receiver side let $\lambda_d$ represent the LLR of dth demapped bit.
\begin{equation}
\begin{split}
\lambda_d=&\mathrm{log}\frac{P(b_d=0|\{V(\mathbf{x})\})}{P(b_d=1|\{V(\mathbf{x})\})}\\
&=\mathrm{log}\frac{\sum_{\mathbf{x}\in \mathcal{X}_d^{(0)}}P(b_d=0|\mathbf{x},\{V(\mathbf{x})\})P(\mathbf{x}|\{V(\mathbf{x})\})}{\sum_{\mathbf{x}\in \mathcal{X}_d^{(1)}}P(b_d=0|\mathbf{x},\{V(\mathbf{x})\})P(\mathbf{x}|\{V(\mathbf{x})\})}\\
&=\mathrm{log}\frac{\sum_{\mathbf{x}\in \mathcal{X}_d^{(0)}}V(\mathbf{x})}{\sum_{\mathbf{x}\in \mathcal{X}_d^{(1)}}V(\mathbf{x})}=\mathrm{log}\frac{\sum_{\mathbf{x}\in \mathcal{X}_d^{(0)}}e^{L(\mathbf{x})}}{\sum_{\mathbf{x}\in \mathcal{X}_d^{(1)}}e^{L(\mathbf{x})}}
\end{split}
\end{equation}
where $\mathcal{X}_d^{(b)}$ denotes the constellation subset with the dth bit being $b\in \{0,1\}$. And then these bit soft message $\{\lambda_d\}$can be further used in channel soft decoding.

\subsection{Low Complexity SCMA Detector}
As mentioned before, the difficulty to deal with the Log-MPA is in formula (6) while calculating exponent operation, we can regard the math problem as to process approximate calculation: $\mathrm{log}\sum\limits_{n=1}^{|com|}e^{C_{jm}^{(n)}}$ and $\lambda_d$
, where Jacobian logarithm \cite{viterbi1998intuitive} is feasible for this issue.
\begin{equation}
\begin{split}
\mathrm{log}(e^{x}+e^{y})=f(x,y)&=max(x,y)+\mathrm{log}(1+e^{-|x-y|})\\
&=max(x,y)+g(|x-y|).
\end{split}\label{log}
\end{equation}
Handling this equation we can calculate the above formula with iterative method
\begin{equation}
\mathrm{log}\sum\limits_{n=1}^{N}e^{x_n}= f(x_n,f(x_{n-1},f(x_{n-2},...,f(x_2,x_1)...))).
\end{equation}
We only need to approximate the second term of the right-hand part of Eq.(\ref{log}) $g(\cdot)$ in the iterative processing, because $max(\cdot)$ operator is easy to implement.
Firstly completely eliminating the second part may be a easy choice, which is elaborated as "Max-Log" $g(z)=0;z\geq 0$ in \cite{zhang2014sparse}

In this work, we use the total least squares criterion, choosing two curve types as prediction model to approximate the function $g(\cdot)$, the first non-linear curve is as follow:
\begin{equation}
 g_1(z)=\left\{
\begin{aligned}
&  \frac{a}{b+z}+c &z\leq d \\
&   0 &z>d\\
\end{aligned}
\right.
\end{equation}
to optimize the total squared error between exact term and $g_1$, we got optimal $a=1.0807$, $ b=1.1657$, $c=-0.1975$, $d=5$ by means of numerical calculation and the mean squared error (MSE) is $5.2246\times10^{-5}$.

\begin{figure}[t]
            \centering
            \includegraphics[width=3.2in,height=2.7in]{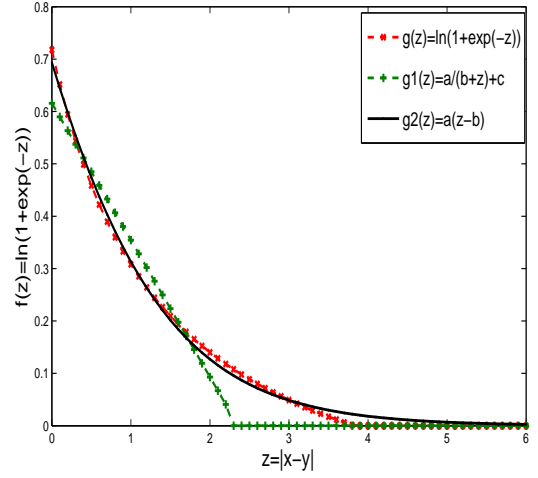}
            \hspace{2in}\parbox{1\linewidth}
{\caption{The value of ln(1+exp(-z))and two approximate curves g1 and g2. }
\label{three_function_curve}}
\end{figure}

The second linear prediction model is given by:\\
\begin{equation}
 g_2(z)=\left\{
\begin{aligned}
& a(z-b) &z\leq b \\
&   0 &z>b\\
\end{aligned}
\right.
\end{equation}
In this model, the optimized parameter is $a=-0.2614$, $b=2.3555$, the MSE between $g_2(\cdot)$and $g(\cdot)$is $5.9254\times10^{-4}$, obviously $g_1$shows  better fitting performance as shown in Fig.\ref{three_function_curve}.

\subsection{Average Mutual Information of SCMA (SCMA-AMI)}
Without loss of generality, we consider the Average Mutual Information (AMI) as a comparison criterion to evaluate the influence of approximation models\cite{xie2012simplified}.

In our sytem model, as bit streams are interleavered and then mapped to SCMA codebooks, on receiver side we can reasonably regard each bit level $b_d$ to a received $\mathbf{y}$ forms an independent channel. For a memoryless channel with perfect channel estimation, it can be assumed that input random variable X is equally taken from the points of the alphabet $\mathcal{X}$, the conventional AMI is defined as \cite{caire1998bit}:
\begin{equation}
\begin{split}
I(X;Y|H)=\sum_{d=1}^DI(B_d;Y|H),
\end{split}
\end{equation}
As we have stated in the previous chapter, SCMA is an overloading multiplexing scheme, in such a case, it is reasonable to define SCMA-AMI as follows:
\begin{equation}
\begin{split}
I_{SCMA}=\mu*I(X;Y|H)=\mu*\sum_{d=1}^DI(B_d;Y|H),
\end{split}
\end{equation}
where $\mu=J/K$, $B_d\in \{0,1\}$ is overloading factor,
\begin{equation}
\begin{split}
I(B_d;Y|H)&=E_{b_d,\mathbf{y},h}[I(b_d;y|h)]\\
&=E_{b_d,\mathbf{y},h}[I(b_d|h)-I(b_d|y,h)]\\
&=1-E_{b_d,\mathbf{y},h}[log_2\frac{p(y,h)}{p(b_d,y,h)}]\\
&=1-E_{b_d,\mathbf{y},h}[log_2\frac{\sum_{\mathbf{x}\in \mathcal{X}}p(y,h,x)}{\sum_{\mathbf{x}\in \mathcal{X}^{(b_d)}}p(b_d,y,h,x)}]\\
&=1-E_{b_d,\mathbf{y},h}[log_2\frac{\sum_{\mathbf{x}\in \mathcal{X}}p(y|h,x)}{\sum_{\mathbf{x}\in \mathcal{X}^{(b_d)}}p(y|h,x)}].
\end{split}
\end{equation}
Theoretically, if we use Maximum a Posteriori demapping criterion (MAP) or Message passing Algorithm (MPA) within cycle-free factor graph, we can derive the SCMA-AMI with formula (18) and (19). For further deduction, we define;
\begin{equation}
L_{d}^{MAP}=\mathrm{log}\frac{P(B_d=0|Y,H)}{P(B_d=1|Y,H)}=\mathrm{log}\frac{\sum_{x\in{\mathcal{X}_d^{(0)}}}p(Y|x,H)}{\sum_{x\in{\mathcal{X}_d^{(1)}}}p(Y|x,H)},
\end{equation}
since
\begin{equation}
\sum_{B_{d}}P(B_d|Y,H)=1.
\end{equation}
Following from (20) and (21), it is trivial to see:
\begin{equation}
\begin{split}
P(B_d=b_d|Y,H)=\frac{e^{L_{d}^{MAP}}(1-b_d)}{1+e^{L_{d}^{MAP}}}=P(B_d|L_{d}^{MAP}).
\end{split}
\end{equation}
Using the definition of conditional mutual information and adding equation(21), (23) into the following formula, we can show that :
\begin{equation}
\begin{split}
&I(B_d;Y|H)=I(B_d;L_{d}^{MAP})=H(B_d)-H(B_d|L_d^{MAP})\\
&=1-\sum_{B_d}\int P(b_d,L_{d}^{MAP})\mathrm{log}\frac{1}{P(b_d|L_{d}^{MAP})}dL_{d}^{MAP},\\
\end{split}
\end{equation}
As mentioned above, we got the theoretical method to obtain SCMA-AMI in formula (19) based on the output of optimal Log-MAP demapper.

 However, actually in SCMA system, the factor graph usually exists cycles, which results in inaccurate posterior output bit LLR $\lambda_d \approx L_d^{MAP}$ and with different simplified algorithm quantity of estimation error will bring out different soft output. In detailed circumstance, we use $I(B_d;\lambda_d)$ to esimate the channel capacity and compare approximate SCMA-AMI between different demapper algorithm.
\begin{equation}
I(B_d;\lambda_d)=h(\lambda_d)-h(B_d|\lambda_d),
\end{equation}

\begin{equation}
h(\lambda_d)=\int p(\lambda_d)\mathrm{log}(\frac{1}{p(\lambda_d)})d\lambda_d,
\end{equation}

\begin{equation}
h(\lambda_d|B_d)=\sum_{B_d}\int p(\lambda_d,b_d)\mathrm{log}(\frac{1}{p(\lambda_d|b_d)})d\lambda_d,
\end{equation}
where$h(\cdot)$ and $h(\cdot|\cdot)$ respectively denote the differential and conditional differential entropy.
To obtain the PDF $p(\lambda_d)$ and $p(\lambda_d|b_d)$, we can use the method of Mathematical Statistics to analyze the sample LLR $\lambda_d$, plot sample histogram to get approximate distribution data\cite{ten2001convergence}. As demonstrated before, we insert formula (24) into formula (25) into (19) to get SCMA-AMI witin specific detector algorithm. In next section, we will give these simulation results.

\begin{figure}
            \centering
            \includegraphics[width=3.2in,height=2.7in]{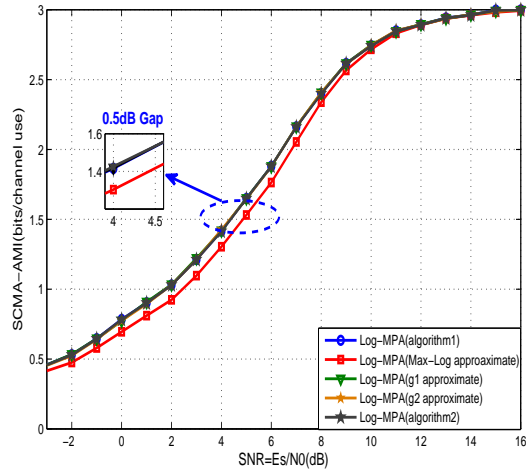}
            \hspace{2in}\parbox{1\linewidth}
{\caption{AMI Comparison of different SCMA detection schemes over AWGN channels }
\label{SCMA_AMI1}}
\end{figure}

\section{Numerical results}
In this section, numerical results are presented to evaluate the performance of the proposed schemes. First, SCMA-AMI associated with different detection algorithms defined in Section III-D are shown in Fig.\ref{SCMA_AMI1} for AWGN channel. This simulation is based on the basic parameters $J=6$, $K=4$, $N=2$, $M=4$, $\mu=J/K$(connection relationship is shown in Fig.\ref{factor_graph}), 6 layers SCMA codebooks are designed according to \cite{taherzadeh2014scma}. The SCMA-AMI loss engendered by simplified structure in algorithm 2 is negligible compared with original MPA in algorithm 1 as well as two approximation methods g1 and g2. Meanwhile, the traditional Max-Log approximation shows obvious information loss especially in low SNR circumstance within almost 0.5dB deficiency.

\begin{figure}[t]
            \centering
            \includegraphics[width=3.2in,height=2.7in]{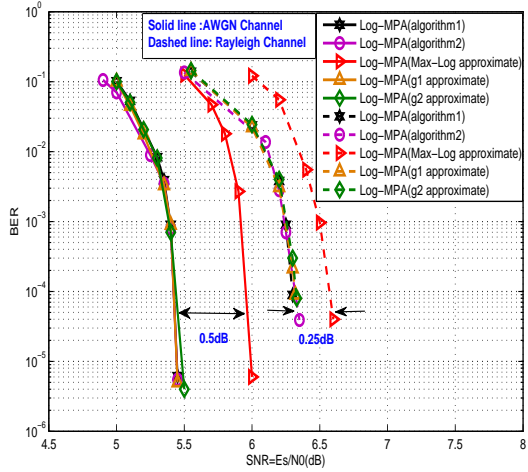}
            \hspace{2in}\parbox{1\linewidth}
{\caption{BER performance of SCMA detection algorithms over AWGN and Rayleigh Fading channels. }
\label{ber_curve2}}
\end{figure}

Second, BER simulation is carried out to evaluate the performance of these above algorithms. The regular systematic LDPC channel coding in China Mobile Multimedia Broadcasting (CMMB) systems is used in both AWGN and Rayleigh fading channel with the code rate $R=1/2$ and 9216 coded bits per block. The channel coefficient $h$ is set to a unit power, i.e., $E(h^2)=1$. It is observed from Fig.\ref{ber_curve2} that in both of two types of channels the proposed three simplified method show almost the same BER performanc, the width of water fall regions are nearly 0.4 dB. Scarcely any losses in BER performance are gernerated by using simplified algorithm1, g1 and g2 approximate methods, while traditional Max-Log method  DPC channel coding presents poor results, within as much as 0.5dB gap over AWGN channel and almost 0.25dB loss over Rayleigh fading channel.

As what we have anticipated, three proposed methods present obvious advantages over traditional Max-Log methods with reasonable calculation complexity to some extent.

\section{Conclusions}
In this paper, we have proposed a simplified SCMA detection algorithm via making prior decisions on some users with higher confidence, which reduces the complexity of further extrinsic information calculation on correlated edges. Further more, based on the logarithm expression derived in this paper, we approximated the iterative formula via curve fitting under the criterion of total least squares. On one hand, to analyze SCMA scheme in theory, the average mutual information of SCMA (SCMA-AMI) has been proposed. Furthermore, the complexity of calculation is significantly reduced while the performance of proposed algorithm considerably outperforms that of  using Max-Log approximation in numerical simulation results. In conclusion, proposed well-performed  algorithms make an excellent tradeoff between the calculation complexity and performance in the SCMA system.

\bibliographystyle{IEEEtran}
\bibliography{xkx_paper}
\end{document}